\documentclass[useAMS,a4paper,fleqn,usenatbib]{mnras}
\usepackage{newtxtext,newtxmath}
\usepackage{mathptmx}
\usepackage{txfonts}
\usepackage[T1]{fontenc}
\usepackage{ae,aecompl}
\usepackage{graphicx}	
\usepackage{amsmath}	
\usepackage{amssymb}	
\usepackage{url}
\usepackage{subfigure}
\usepackage{color}
\usepackage{mathtools}
\newcommand{\sassy}{{\rm SASSy}}

\newcommand{\smurf}{{\sc smurf}}
\newcommand{\cupid}{{\sc cupid}}
\newcommand{\fellwalker}{{\sc fellwalker}}

\DeclarePairedDelimiter\abs{\lvert}{\rvert}

\title[SASSy: a catalogue of beam-sized sources ]{The SCUBA-2 Ambitious Sky Survey: a catalogue of beam-sized sources in the Galactic longitude range 120\degr\ to 140\degr\ }

\author[W. Nettke et al.]{Will Nettke,$^{1}$\thanks{E-mail: willnettke@gmail.com}
Douglas Scott,$^{1}$
Andy G. Gibb,$^{1}$ 
Mark Thompson,$^{2}$\newauthor
Antonio Chrysostomou, $^{3}$
A. Evans,$^{4}$
Tracey Hill,$^{5}$
Tim Jenness,$^{6}$
Gilles Joncas,$^{7}$\newauthor
Toby Moore,$^{8}$
Stephen Serjeant,$^{9}$
James Urquhart,$^{10}$
Mattia Vaccari,$^{11}$\newauthor
Bernd Weferling,$^{12}$
Glenn White,$^{9,13}$
Ming Zhu,$^{14}$
\\
$^{1}$Department of Physics \& Astronomy, University of British Columbia, 6224 Agricultural Road, Vancouver, BC V6T 1Z1, Canada\\
$^{2}$Centre for Astrophysics Research, Science \& Technology Research Institute, University of Hertfordshire, College Lane, Hatfield, \\AL10 9AB, UK\\
$^{3}$Square Kilometre Array Ogranisation, Jodrell Bank Observatory, Lower Withington, Macclesfield, Cheshire, SK11 9DL, UK\\
$^{4}$Astrophysics Group, Keele University, Keele, Staffordshire, ST5 5BG, UK\\
$^{5}$Joint ALMA Observatory, Alonso de Cordova 3107, Vitacura, Chile\\
$^{6}$LSST Project Office, 933 N. Cherry Ave, Tucson, AZ 85721 USA\\
$^{7}$Département de physique, de génie physique et d?optique and Centre de recherche de astrophysique du Québec, \\1045 avenue de la médecine, Université Laval, Québec, Québec, JCMTLSYP 1V 0A6\\
$^{8}$Astrophysics Research Institute, IC2 Liverpool Science Park, 146 Brownlow Hill, Liverpool, L3 5RF, UK\\
$^{9}$Department of Physical Sciences, The Open University, Milton Keynes, MK7 6AA, \\England
$^{10}$Max-Planck-Institut f\"ur Radioastronomie, Auf dem H\"ugel 69, D-53121 Bonn, Germany\\
$^{11}$Astrophysics Group Physics Department University of the Western Cape Private Bag X17, 7535, Bellville, \\Cape Town, South Africa\\
$^{12}$Joint Astronomy Centre, 660 N. A'ohoku Place, University Park, Hilo, Hawaii 96720, U.S.A\\
$^{13}$RAL Space, The Rutherford Appleton Laboratory, Chilton, Didcot OX11 0NL, England\\
$^{14}$National Astronomical Observatories, Chinese Academy of Sciences, 20A Datun Road, Chaoyang District, Beijing, 100012, China\\
}

\date{Accepted XXX. Received YYY; in original form ZZZ}

\pubyear{2016}

\begin{document}
\label{firstpage}
\pagerange{\pageref{firstpage}--\pageref{lastpage}}
\maketitle

\begin{abstract}
The SCUBA-2 Ambitious Sky Survey (SASSy) is composed of shallow 850-$\umu$m
imaging using the Sub-millimetre Common-User Bolometer Array 2 (SCUBA-2) on the
James Clerk Maxwell Telescope.  Here we describe the extraction of a
catalogue of beam-sized sources from a roughly $120\,{\rm deg}^2$ region of the
Galactic plane mapped uniformly (to an rms level of about 40\,mJy), covering
longitude 120\degr\,$<$\,\textit{l}\,$<$\,140\degr\ and latitude
$\abs{\textit{b}}$\,$<$\,2.9\degr.  We used a matched-filtering approach to
increase the signal-to-noise (S/N) ratio in these noisy maps and tested
the efficiency of our extraction procedure through estimates of the
false discovery rate, as well as by adding artificial sources to
the real images.  The primary catalogue contains a total of 189 sources at
850\,$\umu$m, down to a S/N threshold of approximately 4.6. Additionally, we
list 136 sources detected down to ${\rm S/N}=4.3$, but recognise that as we go
lower in S/N, the reliability of the catalogue rapidly diminishes.  We perform
follow-up observations of some of our lower significance sources through small
targeted SCUBA-2 images, and list 265 sources detected in these maps down to
${\rm S/N}=5$.  This illustrates the real power of SASSy: inspecting the
shallow maps for regions of 850-$\umu$m emission and then using deeper targeted
images to efficiently find fainter sources.  We also perform a comparison of
the SASSy sources with the Planck Catalogue of Compact Sources and the
\textit{IRAS} Point Source Catalogue, to determine which sources discovered in
this field might be new, and hence potentially cold regions at an early stage
of star formation.
\end{abstract}


\section{Introduction}\label{sec:intro}

Large-scale astronomical surveys are the drivers for studying the properties of the Universe on both Galactic and extragalactic scales. The SCUBA-2 Ambitious Sky Survey, or \sassy\footnote{Previously known as the SCUBA-2 `All-Sky' Survey.} \citep{2011MNRAS.415.1950M, 2007arXiv0704.3202T} is a James Clerk Maxwell Telescope (JCMT) Legacy Survey (JLS) that seeks to build on results obtained from previous surveys at similar wavelengths, such as with \textit{IRAS} and \textit{Planck}, by mapping large fractions of the sky at wavelengths where the emission is dominated by thermal dust. This paper describes the procedure used to identify beam-sized sources in a region that has now been fully surveyed. The main motivation for this work is to identify new Galactic sources, potentially including some of the youngest regions of early star formation in the outer Galaxy.

The SASSy survey is capable of operating in the poorest weather conditions possible, using a unique scanning design (called pong \citep{PONG}), maximising the potential of SCUBA-2 in the 850-$\umu$m waveband. While SCUBA-2 simultaneously observes at 450\,$\umu$m, those data are considerably noisier for a survey at this depth. This study focuses on a region spanning 120\degr\ to 140\degr\ in Galactic longitude ($\it{l}$) and $-$2.9\degr\ to $+$2.9\degr\ in Galactic latitude (\textit{b}). The JCMT is able to achieve a resolution of 14 arcsec for the 850-$\umu$m channel \citep{2013MNRAS.430.2534D}, which is about 20 times better than the all-sky survey of \textit{Planck} \citep{2016A&A...594A...8P}. 

The maps are made using the Sub-Millimetre User Reduction Facility (\smurf) package \citep{2013MNRAS.430.2545C}, and then matched-filtered to improve the S/N ratio from the relatively noisy raw time-series data of the SCUBA-2 observations. We then extract point-like sources of intense emission from the maps, using a newly implemented source detection algorithm called FellWalker \citep{Berry}. 

We assess the accuracy of our catalogue using completeness tests. We do this by inserting artificial point-like sources with known attributes (such as sky position and peak flux) into a map and attempt to extract them with the same procedure used for finding real sources. Using the FellWalker algorithm, we extract point-like sources from within the maps, and check these results using a false discovery rate (FDR) technique to complement the completeness testing and determine a S/N threshold for sources to be included in the final catalogue. While we recognise that the Galaxy contains sources that are far from point-like, we restrict our analysis to a well defined set of criteria for extracting sources using a matched-filter approach. With anticipation of data collected prior to August 2013 (including a catalogue of regions of extended emission structure) being part of the first release of JCMT Legacy data in March 2016, we leave an investigation of extended sources (including filamentary structures etc.) to a future study.

An important aspect of SASSy is that we carry out an additional follow-up analysis using small targeted maps, with a scanning pattern called daisy, \citep{2014SPIE.9153E..03B}, also obtained using SCUBA-2, to confirm fainter detections listed in the catalogue and develop a list of additional point--sources detected in these observations. This two-level approach makes SASSy considerably more efficient at finding sources than might be expected. The shallow pong maps indicate regions of enhanced emission, in which even very short targeted observations find many individual sources.

We compare the sources detected in the pong extractions with \textit{IRAS} and \textit{Planck} catalogues \citep[as well as a brief comparison with \textit{Akari} data][]{Doi} to determine if any new objects have been detected. At this wavelength of observation we expect any new discoveries to be among the coldest early star forming regions in this field of the Galactic plane. There are of course other related surveys of parts of the Milky Way, such as the Hi-GAL survey carried out by \textit{Herschel} \citep{2010A&A...518L.100M} and ATLASGAL with APEX \citep{2014A&A...565A..75C}, but these typically focus on the inner Milky Way, so there is little or not overlap. Lastly, we look at the 450-$\umu$m data to determine if any sources can be detected at this noisy shorter wavelength, and estimate 450-$\umu$m upper limits for the bulk of the objects in the 850-$\umu$m catalogue.

\section{Observations}\label{sec:observations}

The data were obtained between 2012 May 15 and 2014 May 12 using SCUBA-2 \citep{2013MNRAS.430.2513H} on the 15-m JCMT. Each field was observed for 40 minutes using a 2\,degree rotating pong pattern \citep{2011MNRAS.415.1950M, 2013MNRAS.430.2513H}, scanning at the top speed of 600\,arcsec\,s$^{-1}$, which, as a result of the overscan (i.e.\,the fact that requesting a region of 2\,degrees to be uniformly mapped results in the full extent of the map being somewhat larger), provides reasonably uniform coverage over a roughly circular area nearly 2.5\,degrees in diameter. Pointing was checked regularly using various sources from the JCMT pointing catalogue and found to be consistent to within 3.5\,arcsec throughout the observing period. Further checks were made using sources detected in the SASSy fields and were found to be consistent. The median optical depth at 225\,GHz was 0.14 (ranging from 0.09 to 0.20), yielding a median sensitivity of 180\,mJy at 850\,$\umu$m. The target point-source sensitivity of SASSy is 40\,mJy (after applying a matched-filter), which required four repeats for half of the fields, five or six for most of the remainder, and seven or eight for three fields. The total number of observations is 148, for a total integration time of 98.7 hours.

\begin{figure*}
    \includegraphics[width=\textwidth]{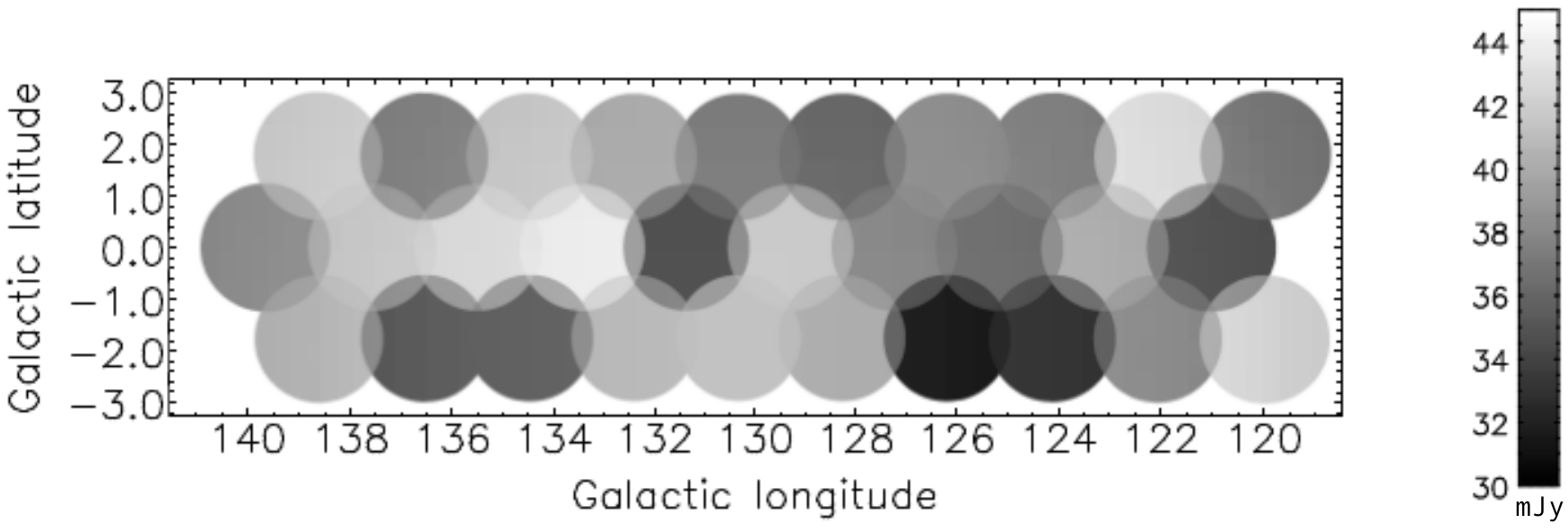}
    \caption{Catalogue coverage depth. The mean rms level in mJy is represented by the greyscale of the individual tiles. There are 30 tiles in all, covering approximately 120\,deg$^2$ in an overlapping, hexagonally-packed arrangement.}
    \label{fig:coverage}
\end{figure*}

The target region for this study, the first contiguous area to be completed from $l$=120\degr\ to $l$=140\degr, $b$=$\pm\,2.9$\degr\,, was sub-divided into fields, which we refer to as `tiles', each 2\,degrees in diameter. Each tile is observed a number of times (until the target sensitivity is reached) adding up to 148 observations in total. The overscan allowed us to use a slightly larger field centre than the initial map diameter, with a spacing of 2.067\,degrees, while preserving some overlap between fields. Three rows of tiles were required to cover the area, offset in $l$ by 1.03\,degrees and in $b$ by 1.79\,degrees, to give a hexagonal close-packed sky coverage. The resulting coverage map (shown in Fig.~\ref{fig:coverage}) is comprised of 30 tiles with a total area of 121\,deg$^2$, and is bounded by 120\degr$<l<$\,140\degr, $-$2.9\degr\,$<b<$\,2.9\degr. The inscribed rectangle covers 19.75\,degrees\,$\times$\,4.87\,degrees (96.2\,deg$^2$).

The daisy observations performed as part of the SASSy programme were carried out on low S/N detections in the pong maps. Each observation uses a pseudo-circular scanning pattern covering a region approximately 12\,arcmin in diameter at 155\,arcsec\,s$^{-1}$. The daisy observations were undertaken with a typical median optical depth at 225\,GHz of 0.13 (ranging between 0.08 and 0.17) giving us an average map rms of approximately 18\,mJy. The 58 supplemental daisy scans included in this study were completed with a total integration time of 5.8 hours.

\section{Data Reduction}\label{sec:datareduction}

\begin{figure}
    \includegraphics[width=\columnwidth]{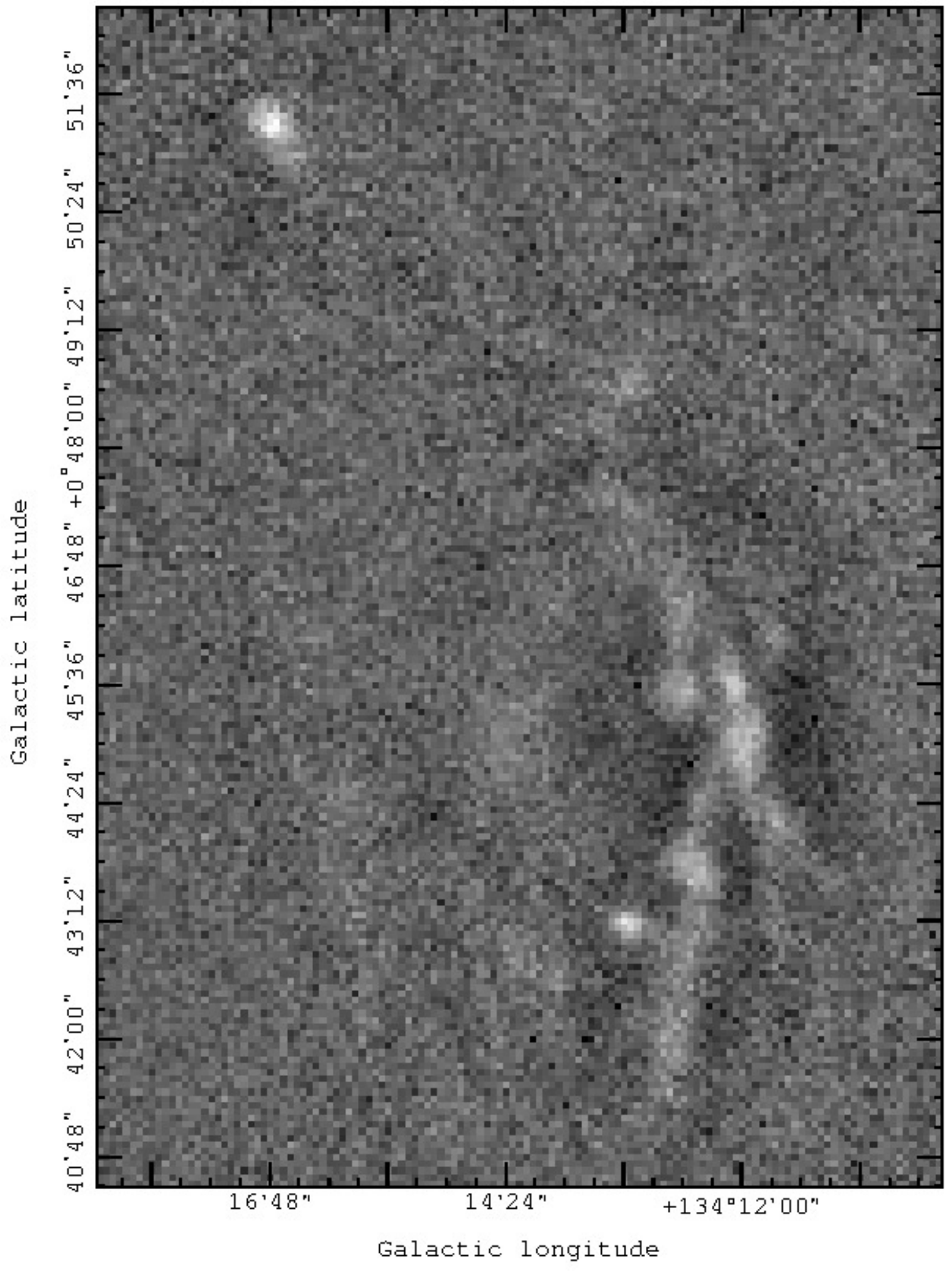}
    \caption{An example of a reduced field, showing part of the W3 star-forming region. As a result of the data reduction process, a negative `bowl' effect can be seen around very bright sources in the map. This particular map shows a portion of the brightest structure in the entire field.}
    \label{fig:reduced_field}
\end{figure}

The data were processed using the \smurf\ iterative map-maker \citep{smurf}, called from the ORAC-DR pipeline \citep{oracdr,2015A&C.....9...40J}. Since the primary goal of SASSy is to detect isolated beam-sized sources, a `blank field' set of configuration parameters was used, which applies a temporal high-pass filter to the data timestreams, corresponding to 200\,arcsec on the sky. This filtering introduces negative `bowls' around strong sources (seen in Fig.~\ref{fig:reduced_field}), but these do not significantly affect the construction of our catalogue because of the spatial filtering approach that we use. There is the potential to miss fainter sources that lie close to bright sources, and extended sources will be potentially broken up into more than one beam-sized source. However, our filtering approach has the advantage of being simple and reproducible.

\begin{figure}
    \includegraphics[width=\columnwidth]{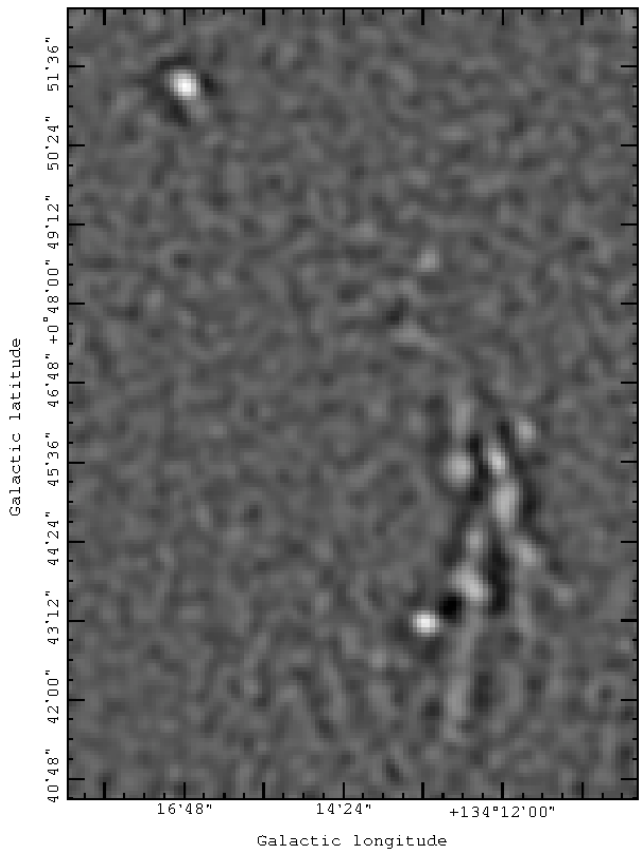}
    \caption{An example of a matched-filtered field. The matched-filter enhances the S/N, but tends to break up larger regions of bright emission into a chain of beam-sized sources, the result of which is flagged in the catalogue for sources with extended emission structure such as this (which is the most extreme example).}
    \label{fig:mf_field}
\end{figure}

Observations of each field were coadded using inverse variance weighting and a `matched-filter' (analogous to the Mexican Hat filter) was applied to the coadd to enhance beam-sized features (seen in Fig.~\ref{fig:mf_field}). The specific implementation adopted was to convolve the map with a model of the JCMT beam \citep{2013MNRAS.430.2534D} and then to effectively subtract a ring on a scale larger than the PSF. To construct this negative rim of the matched-filter, the map was smoothed with a Gaussian having FWHM of 30\,arcsec, and these smoothed maps were subtracted from the originals before the convolution. The coadds were analysed and a S/N map created, on which the source finder was run (described in more detail below).

The data were flux calibrated against the primary calibrators Uranus and Mars, and secondary sources CRL 618 and CRL 2688 \citep[see][]{2013MNRAS.430.2534D}. Due to the filtering in the map-making process, peak fluxes are attenuated by about 16 per cent, an effective conversion factor of 623\,Jy\,beam$^{-1}$\,pW$^{-1}$. We derived this factor in two ways. The first was to process the calibrators using the same method as for the SASSy data. The second approach did the reverse, i.e. SASSy fields with isolated bright beam-sized sources were processed in the same way as the calibrators. Both methods gave the same conversion factor, 2.34, to within two per cent.\footnote{We note that for the March 2016 JCMT data release a slightly higher flux calibration factor of 2.47 has been suggested, however we decide not to apply this correction, since this change lies within the margin of error of source flux densities in our catalogue, described in detail in Section \ref{sec:artificial_testing}.}

While data were obtained simultaneously at 450\,$\umu$m, the atmosphere was too opaque to reach a useful sensitivity for most sources in this band. Given the design of SASSy, it was expected that the 450-$\umu$m data would not be usable, the exception being around the very brightest sources in the W3 region. Only four sources are detected in this part of the surveyed area. For all other 850\,$\umu$m sources we provide an estimate in the catalogue of a 3\,$\sigma$ upper limit at 450\,$\umu$m.

\section{Catalogue}\label{sec:catalogue}

We construct our catalogue using the FellWalker detection algorithm to extract beam-sized sources of emission from within our matched-filtered maps. To test the accuracy of the catalogue and to determine uncertainties in extracted source positions and flux densities, we insert artificial sources into a field (with no bright emission) and attempt to extract them from the map, so that we can assess the completeness. We follow up this test with an estimate of the fraction of false sources recovered, in order to determine a S/N cutoff level for sources to be included in the catalogue. 

\subsection{FellWalker source detection algorithm}
\cupid\ \citep{2007ASPC..376..425B} is a Starlink software package that can be used to extract sources of emission from within our maps. For this catalogue, we specifically adopt the FellWalker algorithm, which has been first implemented for source extraction purposes within the \cupid\, package. The algorithm was developed by David Berry in 2007, and has been well documented and studied for similar purposes of source extraction from other data sets \citep[e.g.][]{Watson,Berry}. Therefore we give only a brief summary here. 

The algorithm has a number of parameters that modify the way in which FellWalker searches for (a process called a `walk'), and identifies, sources. Any pixel above a defined \texttt{Noise} value can be considered as part of a `walk'.  In Table \ref{tab:FellWalker} we give the explicit values used to construct our catalogue that are different from the default parameters (See Section \ref{sec:artificial_testing} for details). From the initial starting point, a `walk' is terminated and a clump of emission is identified as a source provided that it meets the criteria listed in Table \ref{tab:FellWalker}. The basic concept is to identify peaks in the matched-filtered map and to accept these as sources if they are sufficiently peak-like, bright enough, contain enough pixels, and are not too close to other sources.

\begin{table*}
\centering
\caption[]{FellWalker Parameters \citep{sun255,Watson}. The numbers listed here are our chosen parameter values for extracting SASSy sources different from FellWalker defaults, which are optimised for source extraction within our maps, while minimising confusion with noise bumps. Using these values it should be possible to exactly reproduce our catalogue.}
\label{tab:FellWalker}
\begin{tabular}{||  p{0.1\textwidth}  || p{0.7\textwidth} || p{0.1\textwidth}}
\hline
  Parameter Name & Description & Value \\ \hline
  \texttt{AllowEdge} & Defines whether a peak of emission at the edge of the map can be considered part of the output catalogue. The noisiest regions of our maps are nearest to the edge, so we choose not to include sources that touch the edge. & 0 (false) \\ \hline
  \texttt{RMS} & Sets the rms value within the map. We are using S/N maps for the source detection, so we set the rms value to 1. & 1\\ \hline
  \texttt{MinHeight} & Minimum peak flux density a source can have to be included in the catalogue. We choose 4.2 because the lower limit for the supplementary catalogue is 4.3. & 4.2 \\ \hline
  \texttt{MinDip} & Defines the minimum dip (in S/N) between two adjacent peaks in order to be considered part of the same clump of emission. By examining regions within various maps containing closely adjacent sources, we find this value to be most effective for distinguishing peaks between two or more neighbouring sources. & 3 \\ \hline
\hline
\end{tabular}
\end{table*}

\subsection{Artificial source testing and optimisation}\label{sec:artificial_testing}
By inserting a grid of artificial point sources with known locations and peak values into real maps, we can attempt to extract them using the FellWalker algorithm and compare the difference between the input and output characteristics to make error estimates for real sources in the catalogue. To do this we use a SASSy tile in our field cropped to the 2\,degree by 2\,degree region with a mean rms level of 40\,mJy.

To create the grid of artificial sources we use the Kernel Application Package ({\sc kappa}) of the Starlink Software Collection \citep{kappa}. Using the routine \texttt{CREFRAME}, we generate a map of randomly distributed artificial Gaussian-shaped point-like sources with FWHM 14\,arcsec and varying peak value.

\begin{figure*}
    \centering
    \subfigure[Artificial source]
    {
        \includegraphics[width=1.625in]{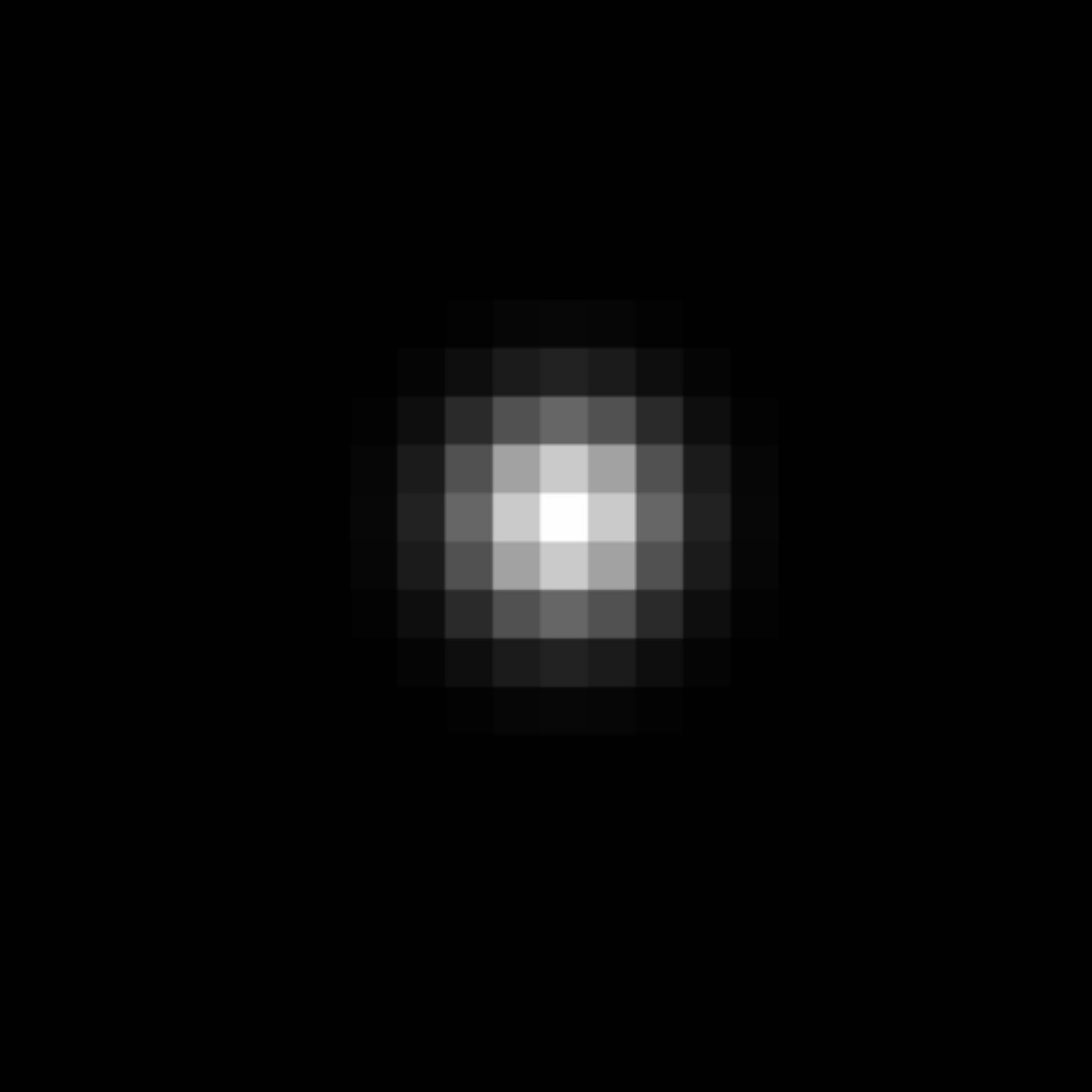}
        \label{fig:artificial_source}
    }
    \subfigure[Reduced]
    {
        \includegraphics[width=1.625in]{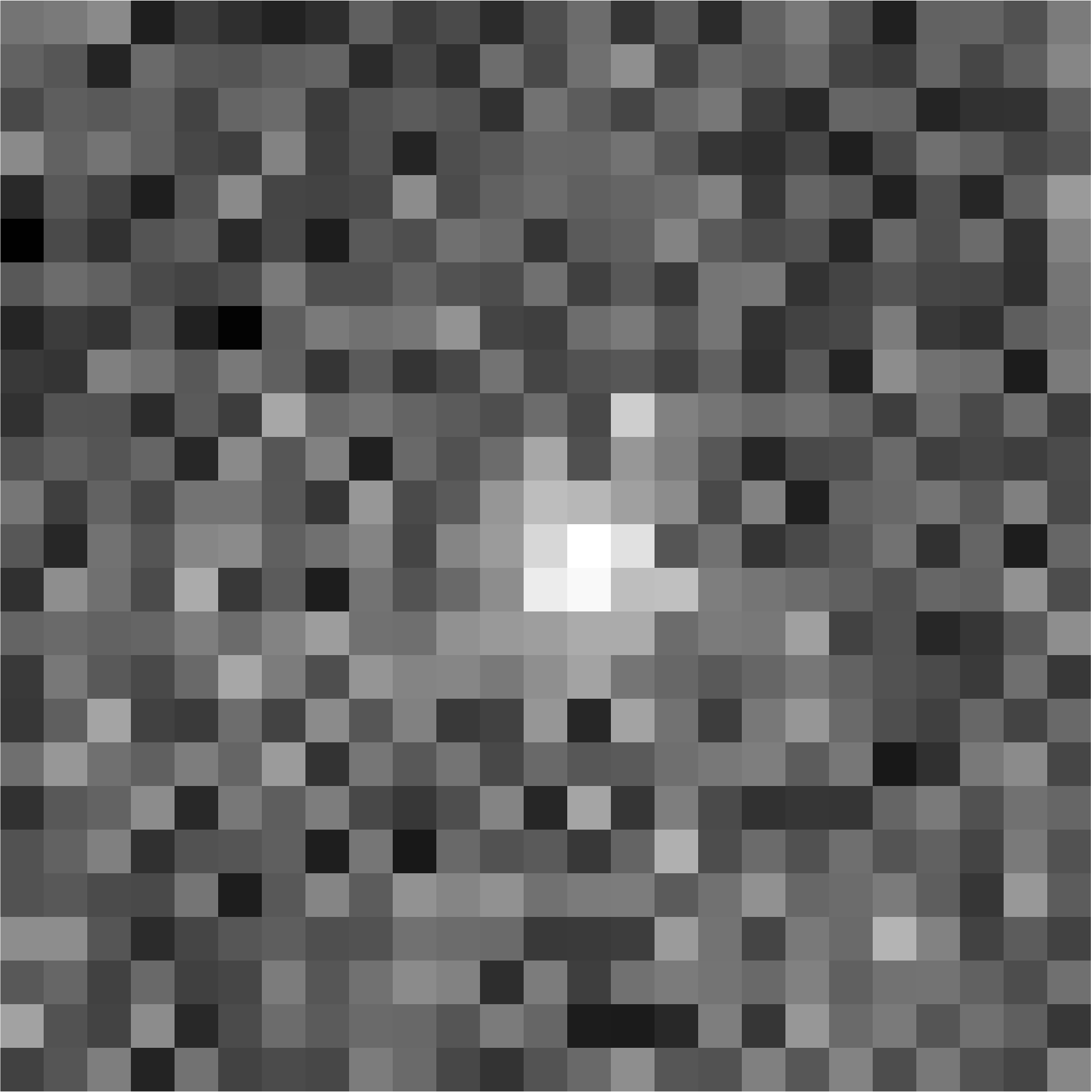}
        \label{fig:reduced}
    }
    \subfigure[Matched-filtered]
    {
        \includegraphics[width=1.625in]{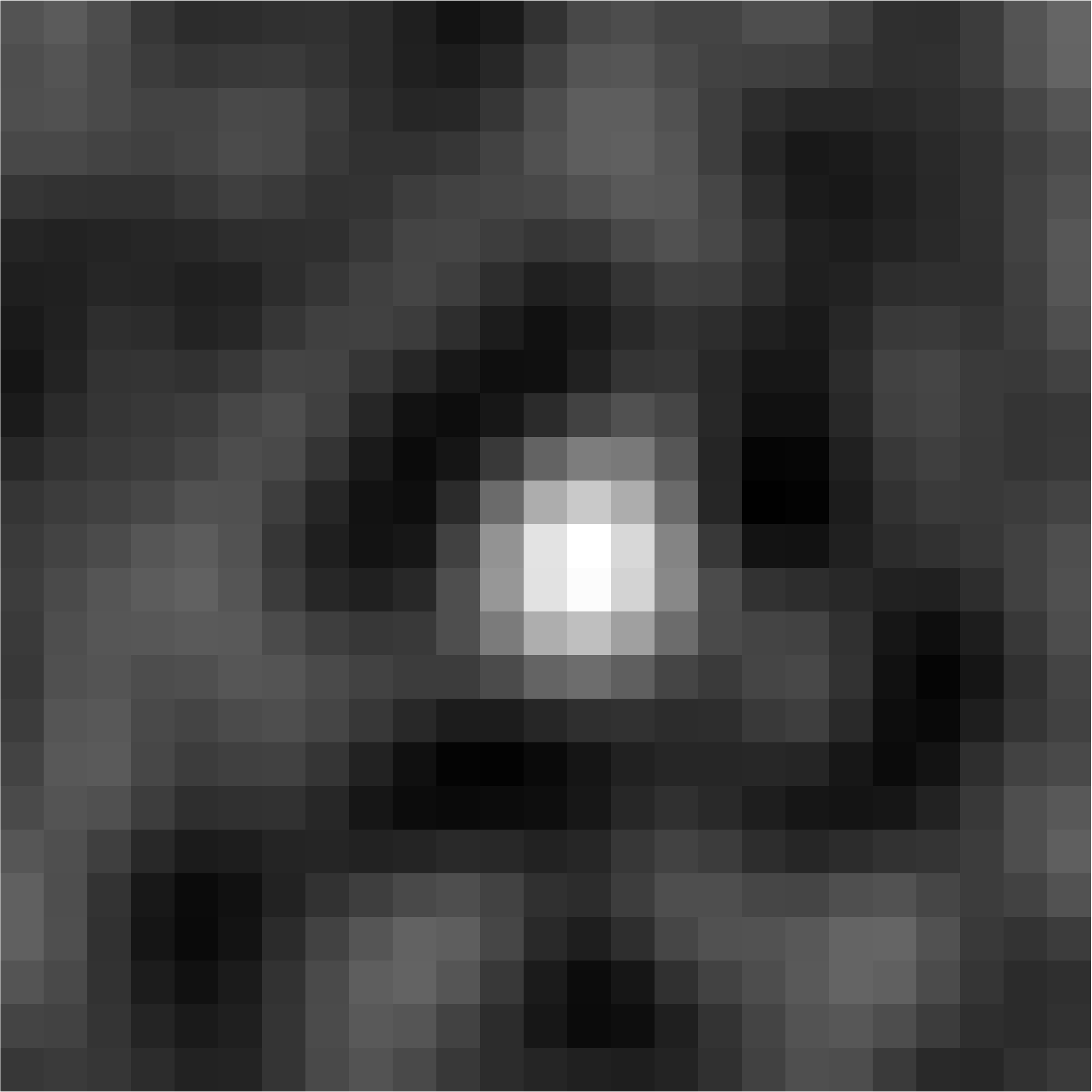}
        \label{fig:matched_filtered}
    }
    \subfigure[Signal-to-noise]
    {
        \includegraphics[width=1.625in]{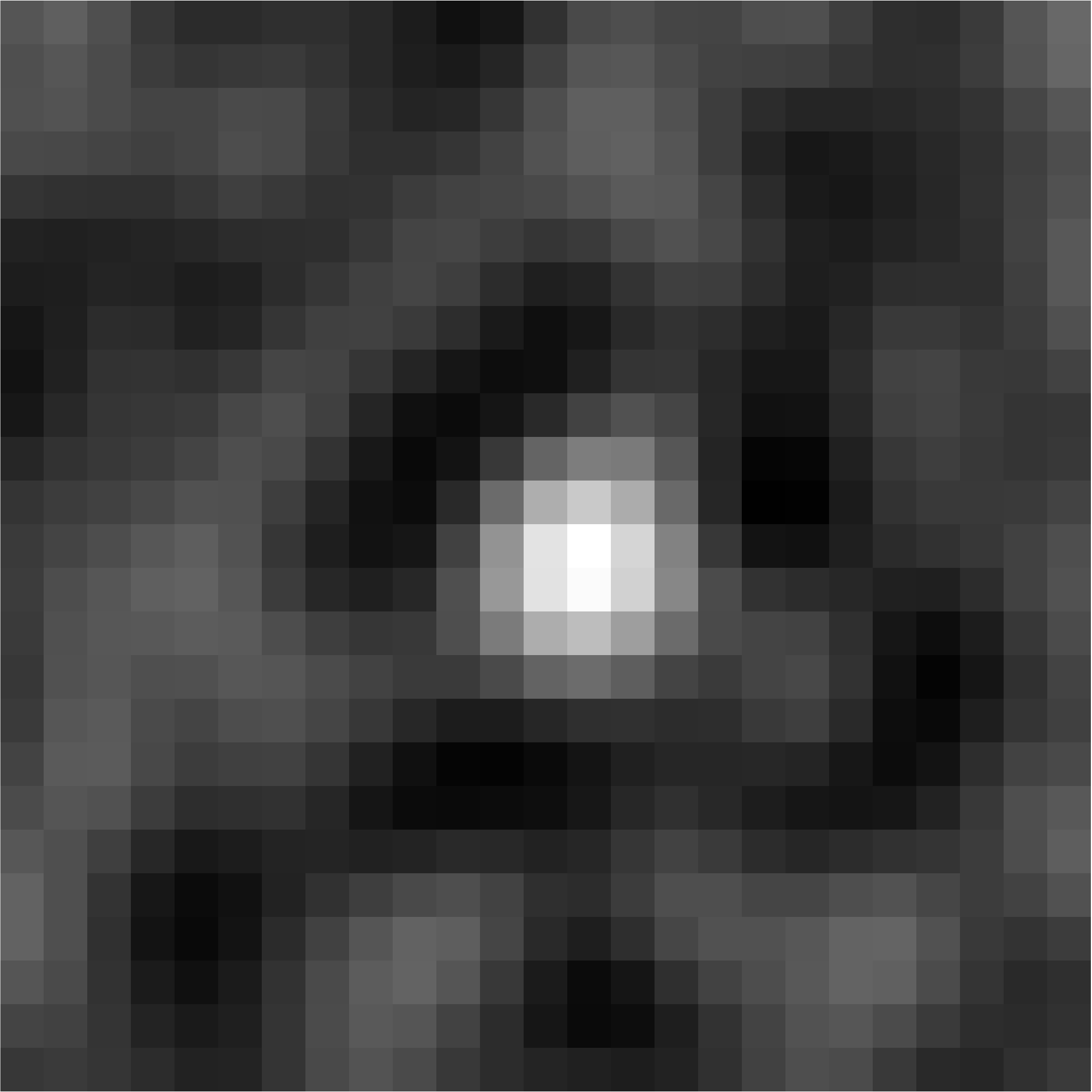}
        \label{fig:signal_to_noise}
    }
    \caption{An artificially generated source as it goes through the data reduction stages. Beginning as a Gaussian point source in panel (a), the source is added during the pipeline reduction stage (therefore including realistic noise) and becomes the image seen in panel (b). From there it undergoes matched-filtering, seen in panel (c), and then a S/N transformation, which removes small variations in the noise and produces a filtered-map in S/N units, seen in panel (d). Panel (d) is almost the same as (c) because the noise is close to uniform across this patch.}
    \label{fig:sources}
\end{figure*}

The sources in the artificial map resemble that shown in Fig.~\ref{fig:artificial_source}, which is representative of the artificial sources inserted. The artificial source insertion process begins at the ORAC-DR pipeline data reduction level. During this stage the artificial sources are inserted into the map once the time-series data have been converted into the appropriate spatial domain, producing sources that look similar to Fig.~\ref{fig:reduced}. We then run our post-processing data reduction on the reduced map, first applying the matched-filter (see Fig.~\ref{fig:matched_filtered}), and then creating a S/N map of that (see Fig.~\ref{fig:signal_to_noise}). We repeat this process two separate times (producing 50 artificial sources in total at each S/N level) on the same tile, the second time adding the fake sources to a rotated map, in order to be sure to spread them among different noise regions.

\begin{figure}
\includegraphics[width=\columnwidth]{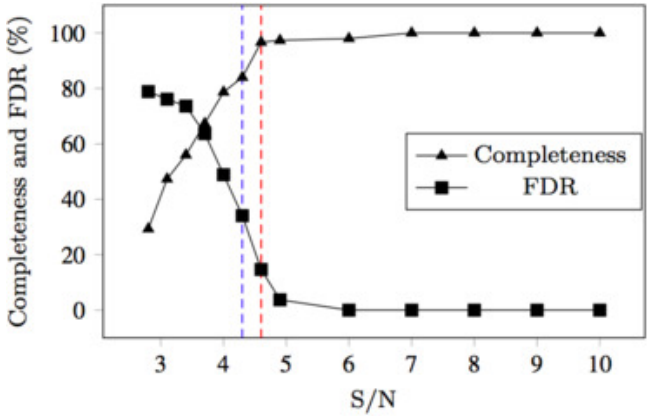}
\caption{Completeness and false discovery rate analysis results. We determine that a compromise between the results of the two tests lies at around $4.6\,\sigma$, represented by the red dashed line. Here the number of recovered sources from the completeness test reaches 96\,\% and the FDR analysis shows approximately 15\,\% false detections. We find this cut-off to be a good compromise between the two tests, and it is supported by the results of the negative source test given in Section \ref{sec:neg_source_test}. The blue dashed line represents the cut-off point chosen for the supplementary catalogue down to 4.3\,$\sigma$ sources; here the completeness reaches 84\,\%, while the FDR estimate yields 34\,\%.}
\label{fig:completeness_fdr}
\end{figure}

\begin{figure}
\includegraphics[width=\columnwidth]{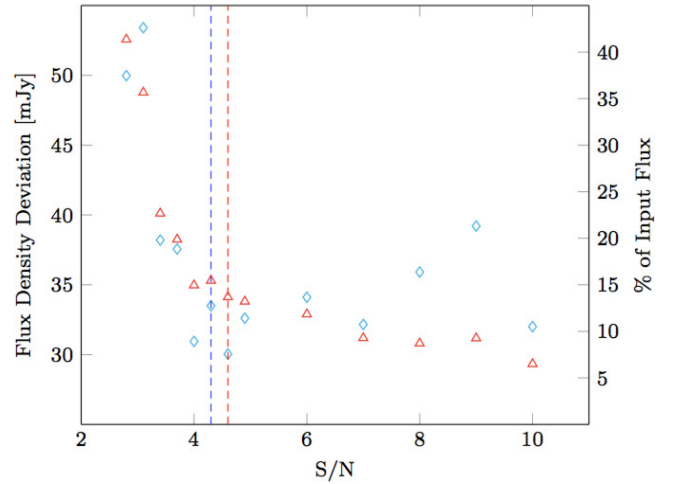}
\caption{Extracted flux density variations. The deviation from the input flux density (shown in the figure as cyan diamonds) falls with S/N, but approaches a value of around 35\,mJy (for this particular tile) at the highest S/N values. The positive bias observed (shown in the figure as red triangles) in the flux density is small enough above $4\,\sigma$ ($<$\,15\,\% of the input flux density) that we have chosen not to correct any source flux densities.}
\label{fig:flux_dispersion}
\end{figure}

\begin{figure}
\includegraphics[width=\columnwidth]{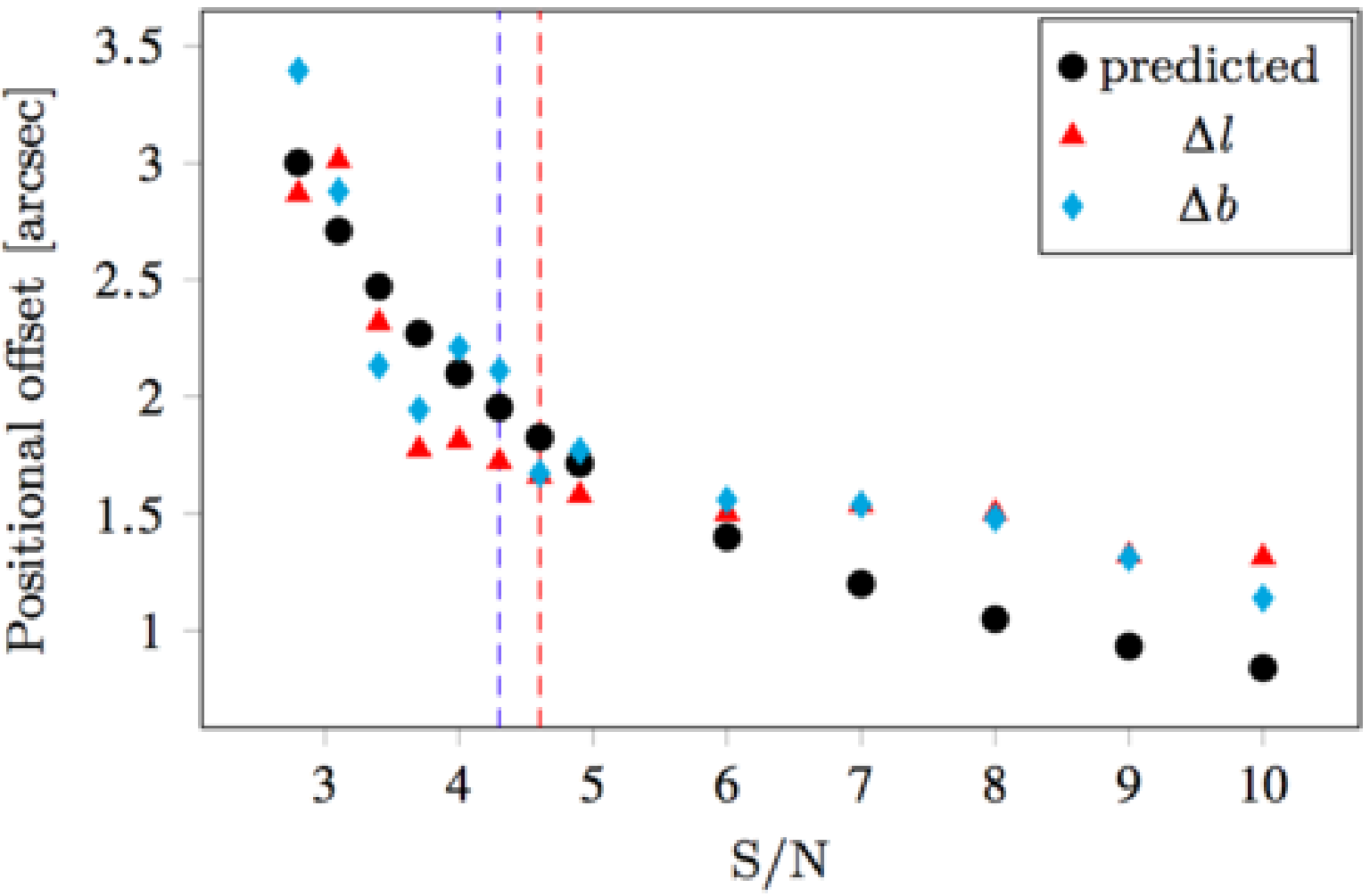}
\caption{Offsets in position for artificial sources. The black circles show the predicted dispersion for the offset in recovered positions (assuming no pointing uncertainties), while the red triangles and blue diamonds show the average measured positional offsets in $\Delta$\textit{l} and $\Delta$\textit{b}, respectively.}
\label{fig:rms_deviation}
\end{figure}

The FellWalker algorithm parameter values we use that are different from the default values, given in Table~\ref{tab:FellWalker}, which represents the optimised parameter values, were determined by exploring the parameter space to ascertain which set of parameters returned the greatest number of realistic sources. We measure the percentage of artificial input sources returned using \fellwalker, which we refer to as the `completeness'; this is shown in Fig.~\ref{fig:completeness_fdr}. We also recover information on positional and flux density uncertainties by comparing the input source characteristics with the output, as shown in Figs.~\ref{fig:flux_dispersion} and \ref{fig:rms_deviation}. For the flux density error we take the rms value of the difference between the input and extracted source parameters at different values of S/N. Additionally we determine the percentage difference in input versus extracted flux, as well as the recovered flux bias. For the positional offset, we use the rms value coming from the difference between the extracted and inserted source positions.

For the error in extracted flux see in Fig.~\ref{fig:flux_dispersion}, we find that the average deviation decreases with S/N, but becomes consistent for the highest S/N values . This result is to be expected, because the noise level is approximately constant across the map. It is also important to note that the source flux density recovered is effectively boosted by the noise contribution in the map. This `flux boosting' effect is a standard result in submm observations, in that given a noisy map with some defined threshold, it is more likely for a detected source to lie on a positive, rather than a negative, noise excursion, resulting in the detected sources being statistically brighter. 

Turning now to positions, we plot the rms offset of recovered positions in Fig.~\ref{fig:rms_deviation}, displaying separate results for offsets in Galactic longitude and latitude. As expected, these fall monotonically with S/N, following the expected relationship, which is that $\Delta\textit{l}$ or $\Delta\textit{b}\,\simeq\,0.6\,\times\,$FWHM$/$(S/N) \citep[see][]{Ivison}.

\subsection{False discovery rate}
We use a false discovery rate (FDR) analysis to complement the results of the completeness testing in order to determine a good compromise for a lower limit on the S/N of sources to be included in the catalogue. Limiting the FDR \citep[described in further detail in][]{Miller}, has been shown to be a robust way of minimising false detections for simulations of sources analogous to those described in this paper.

The FDR is defined as the ratio between the number of false discoveries ($N_{\rm false}$) in a field and the total number of discoveries, both real ($N_{\rm true}$) and false, i.\,e. FDR\,$\equiv\,N_{\rm false}$/($N_{\rm false}$+$N_{\rm true}$).

For the purposes of this study, we use eight different tiles in the entire field to determine the FDR. For the first tile, we construct a `jack-knife' map in the following manner. We coadd scans into two different reduced maps, each with a similar median variance ($\sigma^{2}$), constructed in the same way as described in Section \ref{sec:datareduction}. The difference in $\sigma^{2}$ between the two reduced maps is approximately 5 per cent at most, which is small enough not to affect the statistics of the analysis. Since contributions from any astronomical signal will remain constant between observations (unlike the contribution from sky noise), by taking the difference between the two maps we are left with an effective estimate of the sky noise alone across the tile. We run FellWalker on this `jack-knife' image to estimate $N_{\rm false}$ within the tile for different S/N thresholds. We also determine the number $N_{\rm true}$ using the real map, and thus the FDR for this specific tile, at each specific S/N threshold. We repeat this process using seven other tiles and average the individual FDR results.

Fig.~\ref{fig:completeness_fdr} shows the combined results of the completeness testing and FDR analysis. We find that a reasonable compromise between the two tests lies at $4.6\,\sigma$, where the results of the FDR analysis return approximately 15\,\% false discoveries and the results of the completeness test return 96\,\% of the input sources. We stress that these two tests are calculating different properties of the data, we are simply comparing the results of the two analyses to help determine a reliable balance for the cut-off in S/N for the catalogue. Using $4.6\,\sigma$ as a cutoff for the sources to be included in the catalogue, listed in Table \ref{tab:catalogue}, we find 189 sources in the entire field. While we recognise that a 15\,\% false discovery rate is rather high, we believe it accurately represents the statistics of the catalogue based on the results of the negative source test discussed in Sect.~\ref{sec:neg_source_test}. It is also worth pointing out that similar studies do not often report the FDR, but would be satisfied to use the high completeness level to define the catalogue threshold.

\begin{table*}
	\centering
	\caption{\textbf{Primary Catalogue}: The first column gives the source name, which is the telescope survey and position of the source on the sky in Galactic latitude and longitude. 	The next column gives an ID number for each source, followed by that source's position in both Galactic longitude and latitude, and its Right Ascension and Declination. The final four columns are the extracted S/N value at 850\,$\umu$m, the extracted 850-$\umu$m flux density (measured in mJy), together with a 3\,$\sigma$ upper limit on the flux density at 450\,$\umu$m, and a flag (described in Table~\ref{tab:flags}). The full catalogue can be found online.}
	\label{tab:catalogue}
	\begin{tabular}{|  p{4.5cm}  | p{0.5cm} | p{1.15cm} | p{1.05cm} | p{1.45cm} | p{1.45cm} | p{0.5cm} | p{0.9cm} | p{1.25cm} | p{0.75cm} |}
	\hline
	Source Name         &   ID   &   \textit{l}  &   \textit{b}  &   RA(J2000)  &   Dec(J2000) &   S/N  &   $S_{850}$   &   $S_{450}$   &   Flag    \\ \hline
	[JCMTLSYP LLL.llll$\pm$BB.bbbb]          &       &   [deg] &   [deg] &   [h:m:s] &   [d:m:s] &       &   [mJy]   &       [mJy]   &       \\ \hline
	JCMTLSYP 133.9478$+$1.0643  &   1   &   133.9478    &   $+$1.0643  &   02:27:04.1  &   61:52:23 &   290 &   13800   &       62700   &    E    \\ \hline
	JCMTLSYP 133.7145$+$1.2143  &   2   &   133.7145    &   $+$1.2143  &   02:25:40.0  &   62:05:48 &   170 &   6380    &       33600   &    E    \\ \hline
	JCMTLSYP 133.6945$+$1.2165  &   3   &   133.6945    &   $+$1.2165  &   02:25:31.2  &   62:06:20 &   110 &   3530    &       21300   &    E    \\ \hline
	JCMTLSYP 133.6956$+$1.2098  &   4   &   133.6956    &   $+$1.2098  &   02:25:30.6  &   62:05:58 &   77  &   2590    &       10600   &    E    \\ \hline
	JCMTLSYP 133.7489$+$1.1976  &   5   &   133.7489    &   $+$1.1976  &   02:25:53.8  &   62:04:10 &   69  &   2540    &   $<$ 5800    &        \\ \hline
	\hline
	\end{tabular}
\end{table*}

\begin{figure}
	\includegraphics[width=\columnwidth]{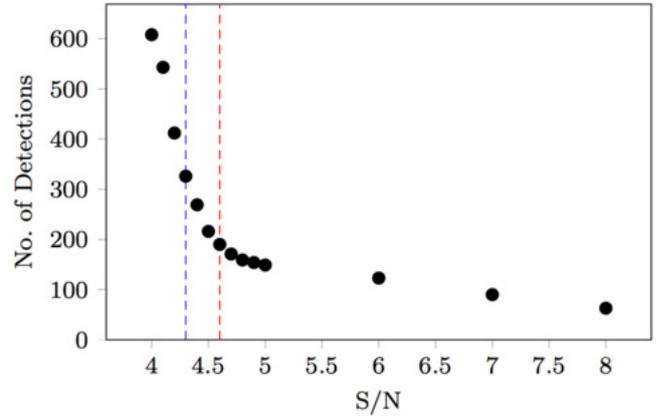}
	\caption{Number of catalogue sources as a function of S/N level. Here we find that the cut-off for the catalogue lies on the shoulder of the distribution (i.e. where the slope changes from steep to shallow). This strengthens our assertion that the S/N cutoff for the catalogue. As the S/N decreases, the number of apparent detections drastically increases, but the opposite happens for the FDR, as shown in Fig.~\ref{fig:completeness_fdr}.}
	\label{fig:num_detections}
\end{figure}

Additionally we find that the number of detections at the cut-off point for the catalogue lies right on the shoulder of the distribution with S/N, as shown in Fig.~\ref{fig:num_detections}. We determine this to be a good point to define the cut-off for the catalogue, because we can include the largest number of sources without compromising the catalogue integrity with too many potential false detections. The 85\,\% level of `real' sources expected here means that our deeper daisy follow-up observations should be efficient for finding sources. We also list in the supplementary catalogue, seen in Table~\ref{tab:sup_catalogue}, sources detected down to a S/N ratio of 4.3, although we emphasize that at this level, where the FDR reaches 34\,\% and the completeness reaches 84\,\% (represented by the blue dashed line in Fig.~\ref{fig:completeness_fdr}), these sources have a non-negligible chance of being noise fluctuations and are only given as a reference for groups interested in follow-up observations. We note that in the case of the supplemental catalogue the FDR cut-off is quite high, 34\%; however the fraction that we expect to be blank or contain no emission is much smaller than this, since many will instead be relatively low S/N peaks (say, 2--4\,$\sigma$). Thus, sources of this S/N level would not be bright enough to be included in our estimates, but more often than not we expect there to be some flux in that position, discussed further in Section~\ref{sec:daisy_pong}. In fact, this is what the daisy follow-up observations confirm empirically, where indeed many of the lower S/N sources (less than 4.6\,$\sigma$, seen in Fig.~\ref{daisy_pong} and online) are observed to have some positive emission at the position of the initial pong measurement.

\begin{table*}
	\centering
	\caption{\textbf{Supplemental Catalogue}: This catalogue lists all sources detected with S/N ranging between 4.3 and 4.5 (rounded to two digits). The column descriptions and catalogue flags are the same as for the primary catalogue (Table~\ref{tab:catalogue}). The full catalogue can be found online.}
	\label{tab:sup_catalogue}
	\begin{tabular}{|  p{4.5cm}  | p{0.5cm} | p{1.15cm} | p{1.05cm} | p{1.45cm} | p{1.45cm} | p{0.5cm} | p{0.9cm} | p{1.25cm} | p{0.75cm} |}
	\hline
	Source Name         &   ID   &   \textit{l}  &   \textit{b}  &   RA(J2000)  &   Dec(J2000) &   S/N  &   $S_{850}$   &   $S_{450}$   &   Flag    \\ \hline
	[JCMTLSYP LLL.llll$\pm$BB.bbbb]          &       &   [deg] &   [deg] &   [h:m:s] &   [d:m:s] &       &   [mJy]   &       [mJy]   &       \\ \hline                                                                                
	JCMTLSYP 133.5745$+$1.1343  &   190 &   133.5745    &   $+$1.1343  &   02:24:18.8  &   62:04:16 &   4.5 &   140 &   $<$ 5800    &       \\ \hline
	JCMTLSYP 132.9815$-$1.9552 &   191 &   132.9815    &   $-$1.9552 &   02:11:27.1  &   59:20:47 &   4.5 &   186 &   $<$ 5300    &      \\ \hline
	JCMTLSYP 130.5390 $+$0.2644  &   192 &   130.5390 &   $+$0.2644  &   01:57:18.2  &   62:09:32 &   4.5 &   162 &   $<$ 4800    &       \\ \hline
	JCMTLSYP 127.1555$+$1.3240   &   193 &   127.1555    &   $+$1.3240   &   01:29:57.1  &   63:52:46 &   4.5 &   276 &   $<$ 5100    &    \\ \hline
	JCMTLSYP 122.8133$+$0.4989  &   194 &   122.8133    &   $+$0.4989  &   00:50:22.6  &   63:22:13 &   4.5 &   173 &   $<$ 5200   &      \\ \hline
	\hline
	\end{tabular}
\end{table*}

\subsection{Negative source test}\label{sec:neg_source_test}

To obtain an alternative estimate of the number of false detections, we perform a negative source test. We begin by inverting each tile in the field, and then run our data processing procedures on these inverted maps, as described in Section~\ref{sec:datareduction}. We determine that the total number of sources contained in this entire inverted field (using \fellwalker) is 27. Given that the total number of sources in the entire field is 189, if we consider 27 of these to be false detections, this means we have a roughly 14\,\% false detection rate from the negative source test. This is entirely consistent with the 15\,\% FDR estimate that we found for our $4.6\,\sigma$ catalogue threshold

\section{Catalogue Attributes}

The catalogue for the entire field studied in this work lists 189 sources in total detected in the pong maps down to a S/N level of 4.6, with an additional 136 sources detected down to a S/N level of 4.3 in our supplementary catalogue. The catalogue, described in Tables~\ref{tab:catalogue} and \ref{tab:sup_catalogue}, gives a number of different attributes for each source detected. These include (in order): source name; identification number; Galactic latitude; Galactic longitude; Right Ascension; Declination; detected S/N at 850\,$\umu$m; detected total flux density at 850\,$\umu$m (in mJy); detected total flux density at 450\,$\umu$m (in mJy, or a 3\,$\sigma$ upper limit for sources detected at 850\,$\umu$m but not at 450\,$\umu$m); and finally a flag column, described in Table~\ref{tab:flags}. 

Table~\ref{tab:flags} describes a flag for sources we believe to be part of some extended emission structure, but are broken into smaller beam-sized sources due to our matched-filter approach. To identify these sources we perform by-eye analysis of regions containing many bright sources close to one another. By looking at the bright emission structure of the reduced field in a particularly populated and bright region (e.g. Fig.~\ref{fig:reduced_field}), it is clear that there are many sources larger than the JCMT beam that would be broken into beam-sized sources by the matched-filter (e.g. Fig.~\ref{fig:mf_field}). We examine many clear examples like this from both the pong and daisy maps and determine an average distance to the next nearest source (44\,arcsec and 38\,arcsec, respectively). We slightly increase these values (by approximately 1 pixel size) to 48\,arcsec for pong sources and 42\,arcsec for daisy sources, in order to ensure we include sources near the higher end of the extended structure criteria. We then flag the sources with an `E' in the catalogue that lie within this distance to the nearest other source. In total we categorise 62 sources in the primary catalogue as extended, along with 2 sources in the supplementary catalogue, and 164 sources in the daisy catalogue. The corresponding number of couples (that is, groups of one or more sources within the limit of being considered extended) is 8, 1, and 24 for the primary, supplemental and daisy catalogues, respectively.

\begin{table*}
\centering
\caption{Description of catalogue flags.}
\label{tab:flags}
\begin{tabular}{||  p{0.1\textwidth}  || p{0.8\textwidth}}
\hline
  Flag Symbol & Description \\ \hline
  E  & Marks if a source is most likely part of a larger structure of emission, broken up by our matched-filter approach. The criterion for a source to be labelled this way is whether or not the nearest source peak lies within 48\,arcsec for pong maps and 42\,arcsec for daisy maps. \\ \hline
  $\sim$P & Marks if a source was not matched in the \textit{Planck} catalogue. \\ \hline
  $\sim$I & Marks if a source was not matched in the \textit{IRAS} catalogue. \\ \hline
  D & Marks if a source that was followed up by a daisy observation was confirmed to be real. \\ \hline
  $\sim$D & Marks if a source was undetected in a follow-up daisy observation. \\ \hline
\hline
\end{tabular}
\end{table*}

\section{Follow-up using daisy maps}\label{sec:daisy_pong}

\begin{figure*}
\includegraphics[width=\textwidth]{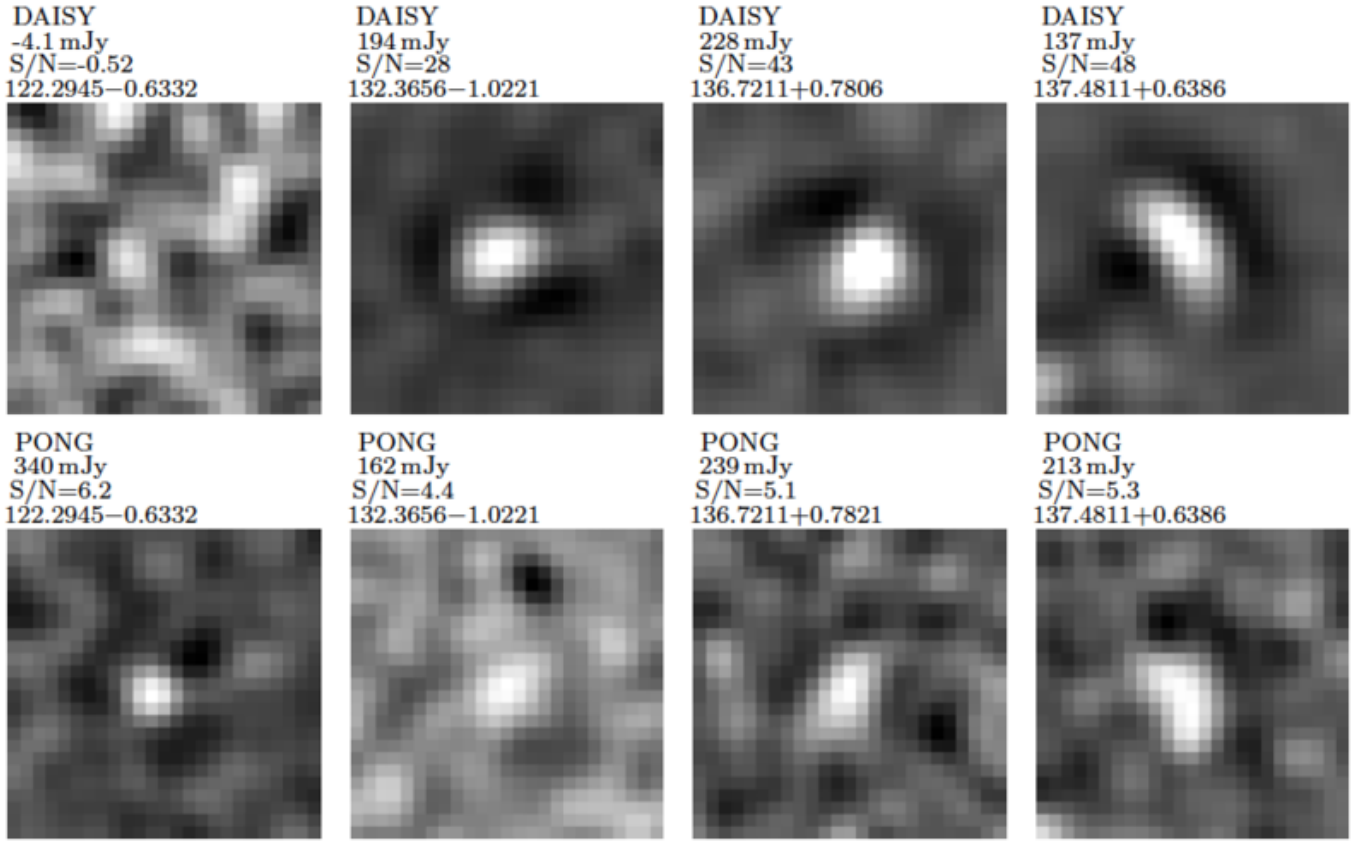}
\caption{Daisy (top row) and pong (bottom row) map source comparison for some representative examples. The leftmost image shows an apparent detection in the pong map that is not recovered in the follow-up daisy observation. The other three image pairs show relatively low S/N sources in the pong observations that \textit{are} recovered at high S/N in the daisy observations. The full set of images can be found online.}
\label{daisy_pong}
\end{figure*}

We performed follow-up observations on 58 sources, which include many of the fainter sources found in the field using a daisy scanning pattern. These daisy maps are much smaller, targeted follow-up observations, carried out on relatively low S/N detections in the pong fields. We use these maps as additional observations to confirm detections in the SASSy field and to find further sources at fainter flux densities. daisy maps are generated using a pseudo-circular scanning procedure, scanning a field approximately 12\,arcmin in diameter at 155 arcsec s$^{-1}$. Thus they are very fast to complete and allow us to achieve a typical rms of 18\,mJy over a small area, giving us a much better S/N ratio than in the pong maps. The daisy observations do not correspond to a specific range of S/N in the final catalogue. Instead, these follow-up observations were performed on sources detected in SASSy with preliminary estimates of S/N, and in some cases prior to the recovery of a reasonable rms level of around 40\,mJy in a pong map. The decision to make the follow-up observations was taken early on in the survey and were not always based on S/N estimates that are the same as those in the final pong-based catalogue. Additionally, some of the follow-up observations depicted in Fig.~\ref{daisy_pong} (with additional examples found online), did not meet the criterion to be included in the final pong catalogue for reasons such as insufficient peak flux or source size.

At the time of this writing, 58 daisy follow-up observations were taken in the catalogue field. For the purpose of consistency, the data reduction methods and source extraction parameters are the same as those used for the pong data taken in the SASSy tiles.

An example comparison between daisy and pong maps of different S/N sources is shown in Fig.~\ref{daisy_pong}. In these representative cases, we see that three of the pong sources show up as real emission in the daisy maps, while one appears to be noise. From the pong catalogue, we determine that only two sources followed up with daisy observations appear to be genuinely false detections; these are marked in the final catalogue. 

We did not perform FDR or completeness procedures on the daisy maps. Instead, to obtain a rough sense of the number of `false discoveries' in the daisy maps, we examine 30 sources discovered in the daisy maps between 4 and 5\,$\sigma$. When we look at the absolute difference in flux at the peak of the source in the daisy map with the flux recovered at that same location in the pong map, we see a clear indication of `flux boosting' in the pong source extraction, with a mean difference of 92\,mJy and median of 74\,mJy. For the purposes of this study we choose not to correct for this effect, but at least mention the estimates found in this test, to give an empirical determination of the strength of this `flux boosting'. Additionally, we see that only 3 of the 30 sources have a flux less than 1$\sigma$ in the pong maps (corresponding to a 16 per cent chance these sources are pure noise). This suggests that only a small fraction of the pong sources are genuinely `false discoveries', in the sense of just being noise fluctuations. So although many of them have been flux-boosted, and they are not necessarily detected in the pong maps, they nevertheless still generally have positive flux, albeit at a lower level than the recovery threshold in the FDR tests. We repeat these tests by broadening our search radius out to 2.5\,pixels (approximately 10\,arcsec) around the location of the daisy source, looking for the largest peak in the pong map in this search area. Here we expect to see a positive bias between the two flux estimates, and indeed 30 out of 30 sources are recovered with positive flux greater than 1$\sigma$ in the pong maps. The `flux boosting' factor in this case has a mean difference of 59\,mJy and median of 57\,mJy.

\begin{table*}
\centering
\caption{\textbf{Daisy Catalogue}: This catalogue lists the Source Name, ID, Galactic longitude and latitude, Right Ascension, Declination, extracted S/N at 850\,$\umu$m, extracted 850-$\umu$m flux density (measured in mJy), and a flag (E) indicating if the source appears to be part of a larger emission structure that was broken up by the matched-filter. The full catalogue can be found online.}
\label{tab:daisy_catalogue}
\begin{tabular}{|  p{4.5cm}  | p{0.7cm} | p{1.15cm} | p{1.05cm} | p{1.45cm} | p{1.25cm} | p{0.5cm} | p{0.9cm} | p{1cm} |}
\hline
Source Name         &   ID   &   \textit{l}  &   \textit{b}  &   RA(J2000)  &   Dec(J2000) &   S/N  &   $S_{850}$   &   Flag     \\ \hline
[JCMTLSYP LLL.llll$\pm$BB.bbbb]          &       &   [deg] &   [deg] &   [h:m:s] &   [d:m:s] &     &   [mJy]    &  \\ \hline
JCMTLSYP 133.9479 $+$1.0645	&	1D	&	133.9479	&	$+$1.0645	&	02:27:04.1	&	61:52:23	&	500	&	11400	&	E	\\ \hline
JCMTLSYP 121.2965 $+$0.6572	&	2D	&	121.2965	&	$+$0.6572	&	00:36:46.5	&	63:28:55	&	210	&	2200	&	E	\\ \hline
JCMTLSYP 133.7156 $+$1.2156	&	3D	&	133.7156	&	$+$1.2156	&	02:25:41.3	&	62:05:53	&	210	&	5050	&	E	\\ \hline
JCMTLSYP 126.7142 $-$0.8210	&	4D	&	126.7142	&	$-$0.8210	&	01:23:32.8	&	61:48:53	&	140	&	885	&	E	\\ \hline
JCMTLSYP 133.6956 $+$1.2167	&	5D	&	133.6956	&	$+$1.2167	&	02:25:31.9	&	62:06:22	&	140	&	2980	&	E	\\ \hline
\hline
\end{tabular}
\end{table*}

\begin{figure*}
    \includegraphics[width=\textwidth]{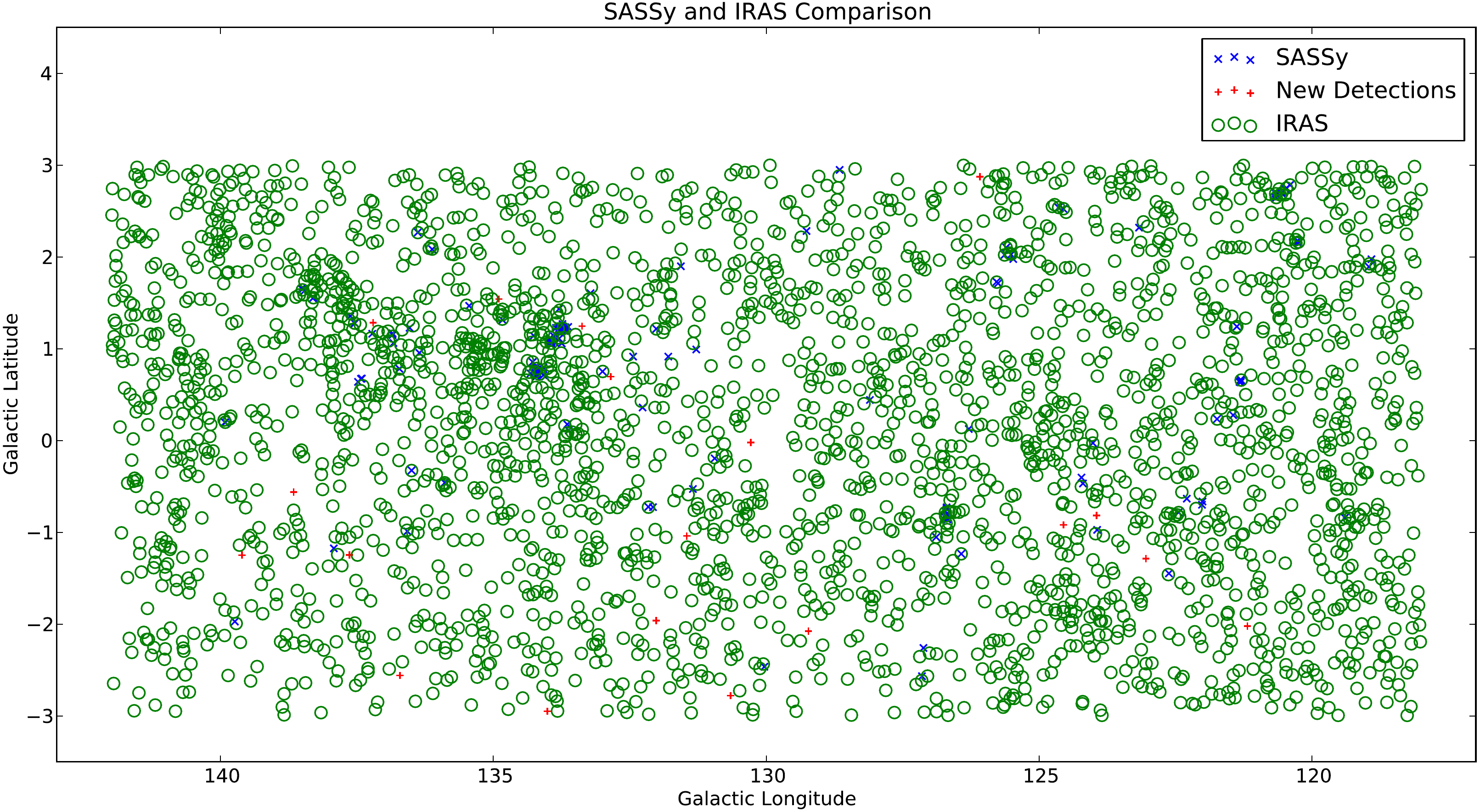}
    \caption{Comparison between the SASSy catalogue and the \textit{IRAS} Point Source Catalogue. The green circles represent (to scale) 1.5 times the \textit{IRAS} beam size (i.e.\,7.05\,arcmin). These circles are centred on each \textit{IRAS} detection (2762 in total), and if a source detected in the SASSy catalogue falls outside these circles it is marked as a previously undetected source (represented as \textcolor{red}{+}). If a source in the SASSy catalogue falls within a circle it is considered to be previously contained in the \textit{IRAS} survey and is marked by \textcolor{blue}{$\times$}. We find 19 new detections, as listed in the primary catalogue.}
    \label{fig:sassy_iras}
\end{figure*}

\begin{figure*}
    \includegraphics[width=\textwidth]{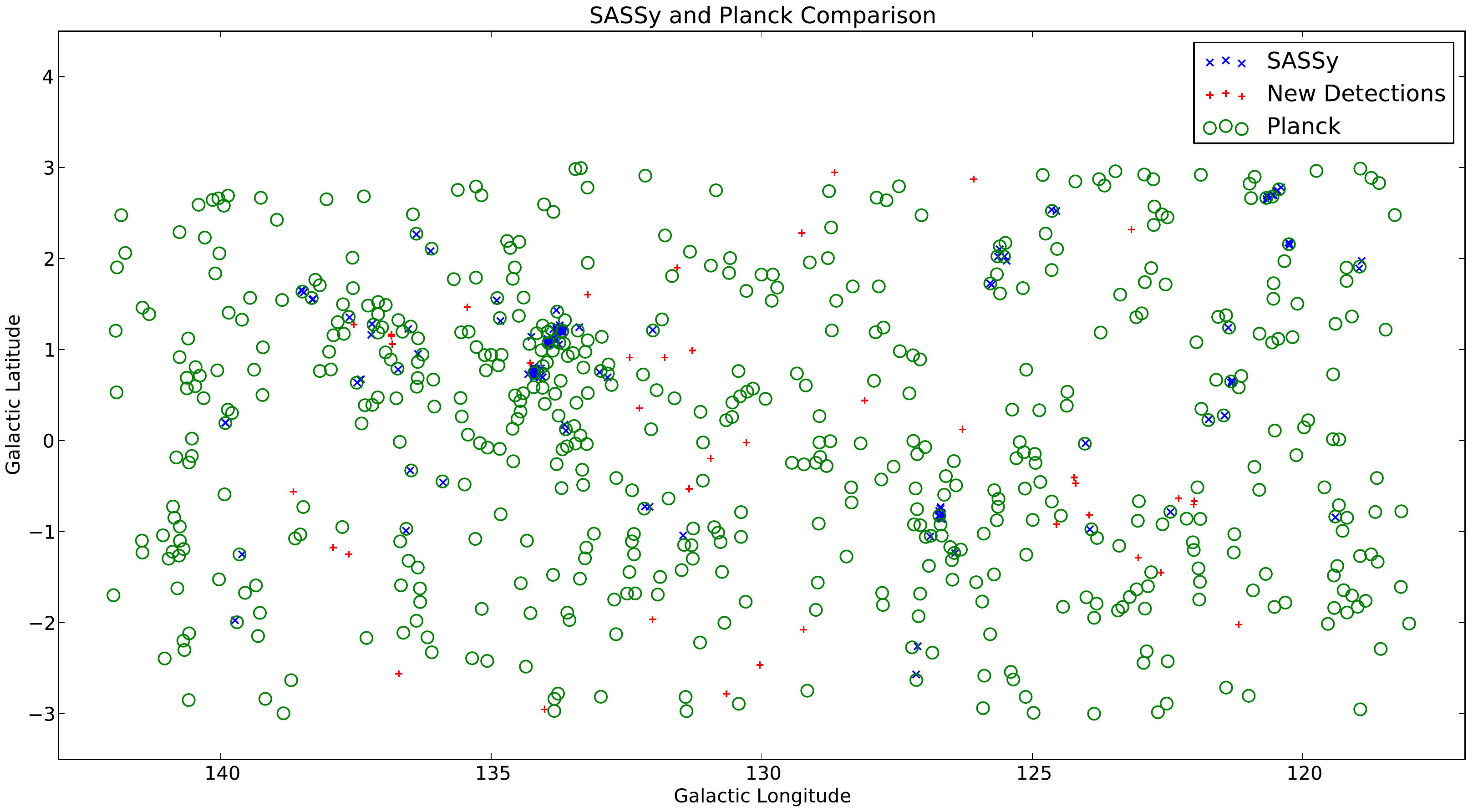}
    \caption{Comparison between the SASSy catalogue and the Planck Catalogue of Compact Sources and Planck Catalogue of Galactic Cold Clumps. The green circles represent (to scale) 1.5 times the \textit{Planck} beam size (i.e.\,7.5\,arcmin). These circles are centred on each \textit{Planck} detection (545 in total after the omission of duplicate sources between PGCC and PCCS), and if a source detected in the SASSy catalogue falls outside these circles it is marked as a previously undetected source (represented as \textcolor{red}{+}). If a source in the SASSy catalogue falls within a circle it is considered to be previously contained in the \textit{Planck} survey and is marked by \textcolor{blue}{$\times$}. We find 41 new detections in our primary catalogue.}
    \label{fig:sassy_planck}
\end{figure*}

While the target for each daisy observation was a relatively low S/N source, the 12\,arcmin map diameter allowed us to go deeper in regions with bright emission close by, which is why the majority of daisy sources in the daisy supplemental catalogue (see Table~\ref{tab:daisy_catalogue}) are a part of an extended emission structure that we mark with an `E' flag. Due to the relatively better depth of the daisy maps compared with the pong maps, and the fact that these follow-up observations are being done with some sense of where a large number of sources may be (in a field we initially measure with pong observations), we find a relatively large number of sources in these maps, even though they are much smaller. The daisy supplemental catalogue provides 265 sources above the 5$\sigma$ level obtained in only 5.8 hours of relatively poor weather; this is a very effective way of finding fainter sources. Since $>$\,200 sources were found in only about 10\,\% of the observing time, the combination of pong surveys and daisy follow-up increases the rate of finding sources by a factor of approximately 20. The decision to set the threshold at 5$\sigma$ was made to ensure that the sources detected in the daisy observations are highly reliable. We conclude that by surveying large regions with pong-style scanning patterns to isolate very bright emission, and following up these observations with smaller targeted daisy observations around these bright areas to go deeper, is an effective way to resolve fainter structures close to these bright spots.

\section{Comparison with other catalogues}

One of the primary motivations for SASSy was to find new cold Galactic sources. Although a full investigation of the spectral energy distributions (SED) of the objects in our catalogue will be left to future investigations, it is nevertheless worth carrying out a simple cross-match with existing catalogues. We perform a preliminary comparison of our main catalogue with \textit{IRAS} \citep{1988iras} and \textit{Planck} \citep{2016A&A...594A..26P,2016A&A...594A..28P} catalogues, obtained through the IPAC catalogue service, to match at other wavelengths and to determine how many new sources may have been discovered.

For comparison with the \textit{IRAS} Point Source Catalogue (PSC), we looked at the sources found in the 100-$\umu$m band, between 118\,$\degr$\,$<$\,\textit{l}\,$<$\,142\,$\degr$ and $-$3\,$\degr$\,$<$\,\textit{b}\,$<$\,3\,$\degr$, to determine if any new cold sources had been discovered. In this field, shown in Fig.~\ref{fig:sassy_iras}, the \textit{IRAS} PSC contains 2762 sources in total, with a limiting flux density of 1535\,mJy for a beam size of 4.7\,arcmin, with varying levels of significance. Since we are mainly interested in sources that are likely detected at 850\,$\umu$m, we impose a limit on the \textit{IRAS} sources in this field to have $\mathit{S}$(100\,$\umu$m)\,$>$\,$\mathit{S}$(60\,$\umu$m).

For comparison with \textit{Planck}, we looked at the sources found in the 353-GHz (approximately 850-$\umu$m) band in the same region for both the compact source catalogue (PCCS) and galactic cold clump catalogue (PGCC), shown in Fig.~\ref{fig:sassy_planck}. These catalogues contain 659 sources in total (459 in the PCCS and 200 in the PGCC) in this range down to a limiting flux density of 727\,mJy for a beam size of 5\,arcmin. We determine that 114 of the 659 sources in these two catalogues are overlapping by conservatively defining an overlapping source as a source in one catalogue that is within a distance of 5\,arcmin (the \textit{Planck} beam width) of at least one other source in the other catalogue. We omit each duplicate source in Fig.~\ref{fig:sassy_planck}, representing 545 sources in total from the \textit{Planck} catalogues. Although SASSy does not go much deeper than the \textit{Planck} all-sky survey, it is important to realise that the \textit{Planck} maps are quite confused in the Galactic plane, because of the much larger beam. Hence SASSy should be able to do significantly better than the \textit{Planck} catalogue in confused regions, and of course our daisy maps are substantially deeper than \textit{Planck}. 

We test our catalogue against each of these other surveys by looking in a region around sources out to 1.5\,$\times$\,FWHM of their beams (4.7 and 5\,arcmin for \textit{IRAS} and \textit{Planck}, respectively). This search radius is chosen in order to contain essentially all true matches, while still representing a tiny fraction of the total image area. We find 19 new discoveries in comparison to \textit{IRAS} and 41 new sources in comparison with \textit{Planck} in this field for our main catalogue, and determine that 13 sources were previously undetected by either the \textit{IRAS} PSC, or the \textit{Planck} PCCS/PGCC. We note that a brief examination of 90-$\umu$m \textit{Akari} data \citep{Doi} agrees with our conclusions drawn from a comparison with the \textit{IRAS} catalogue; for the 19 sources we find that are not recovered in \textit{IRAS}, we determine the same to be true for \textit{Akari}.

\section{Conclusions}

We have reduced SASSy data from the field bounded by $120^{\circ} < l < 140^{\circ}$, $-2.9^{\circ} < b < 2.9^{\circ}$, using the ORAC-DR pipeline and a matched-filtering approach, to achieve a typical rms level of 40\,mJy for beam-sized source detection. Using the FellWalker, source extraction algorithm we determine there to be 189 sources within this field above a threshold of 4.6\,$\sigma$. We list these in Table~\ref{tab:catalogue}, and also include the positions of an additional set 136 sources down to a S/N of 4.3.

The catalogue threshold was determined through the combined use of a false discovery rate estimation and a completeness test using artificial sources. The results of the completeness test also allowed us to estimate the positional offsets and flux density uncertainties. Results from a negative source test, as well as looking at the relationship between the number of detections and S/N level, confirm that the threshold chosen for the catalogue is a reasonable compromise between completeness and false discovery rate.

We confirm many of our sources using small, targeted SCUBA-2 observations, thus verifying the robustness of our catalogue. These observations also enable us to construct a deeper catalogue of sources in regions of bright emission. This conclusion identifies the real scientific power of SASSy: we use shallow, wide swath observations to identify regions of emission by utilizing the matched-filter approach; then we perform quick follow-up observations that go much deeper but only cover a smaller field centred around the faint source. This allows us to uncover more structure around the original sources, giving a very efficient process for finding fainter sources.

Lastly, we perform a comparison of our main catalogue with the \textit{IRAS} and \textit{Planck} catalogues in this field. We confirm many regions of faint emission in our catalogue, but also identify 19 new detections in comparison with \textit{IRAS}, and 41 new detections in comparison with \textit{Planck}, 13 of which were found in neither catalogue.

\section{Acknowledgements}

This research was supported by the Natural Sciences and Engineering Research Council (NSERC) of Canada. The James Clerk Maxwell Telescope has historically been operated by the Joint Astronomy Centre on behalf of the Science and Technology Facilities Council of the United Kingdom, the National Research Council of Canada and the Netherlands Organisation for Scientific Research. Additional funds for the construction of SCUBA-2 were provided by the Canada Foundation for Innovation. The Starlink software \citep{2014ASPC..485..391C} is currently supported by the East Asian Observatory. We are grateful to the JCMT staff and numerous observers for helping to take data for the SASSy project and we thank Gaelen Marsden for helpful discussions. This research has made use of the NASA/IPAC Infrared Science Archive, which is operated by the Jet Propulsion Laboratory, California Institute of Technology, under contract with the National Aeronautics and Space Administration. 

\section{Supplementary Information}

This work provides supplementary information that can be found online in full. Three additional tables are included online to complete the examples given in Tables~\ref{tab:catalogue}, \ref{tab:sup_catalogue}, and \ref{tab:daisy_catalogue}. There is also a figure that can be found online to complete the example of the daisy and pong source comparison given in Figure~\ref{daisy_pong}. A ReadMe file is also included which provides some additional information for the files mentioned above.

\bibliography{sassy_beam_sized_source_catalogue}

\begin{thebibliography}{}
\makeatletter
\relax
\def\mn@urlcharsother{\let\do\@makeother \do\$\do\&\do\#\do\^\do\_\do\%\do\~}
\def\mn@doi{\begingroup\mn@urlcharsother \@ifnextchar [ {\mn@doi@}
  {\mn@doi@[]}}
\def\mn@doi@[#1]#2{\def\@tempa{#1}\ifx\@tempa\@empty \href
  {http://dx.doi.org/#2} {doi:#2}\else \href {http://dx.doi.org/#2} {#1}\fi
  \endgroup}
\def\mn@eprint#1#2{\mn@eprint@#1:#2::\@nil}
\def\mn@eprint@arXiv#1{\href {http://arxiv.org/abs/#1} {{\tt arXiv:#1}}}
\def\mn@eprint@dblp#1{\href {http://dblp.uni-trier.de/rec/bibtex/#1.xml}
  {dblp:#1}}
\def\mn@eprint@#1:#2:#3:#4\@nil{\def\@tempa {#1}\def\@tempb {#2}\def\@tempc
  {#3}\ifx \@tempc \@empty \let \@tempc \@tempb \let \@tempb \@tempa \fi \ifx
  \@tempb \@empty \def\@tempb {arXiv}\fi \@ifundefined
  {mn@eprint@\@tempb}{\@tempb:\@tempc}{\expandafter \expandafter \csname
  mn@eprint@\@tempb\endcsname \expandafter{\@tempc}}}

\bibitem[\protect\citeauthoryear{{Beichman}, {Neugebauer}, {Habing}, {Clegg}
  \& {Chester}}{{Beichman} et~al.}{1988}]{1988iras}
{Beichman} C.~A.,  {Neugebauer} G.,  {Habing} H.~J.,  {Clegg} P.~E.,
  {Chester} T.~J.,  eds, 1988, {Infrared astronomical satellite (IRAS) catalogs
  and atlases. Volume 1: Explanatory supplement}  Vol. 1

\bibitem[\protect\citeauthoryear{Berry}{Berry}{2013}]{sun255}
Berry D.~S.,  2013, Starlink User Note 255.1.
2.0.0 edn

\bibitem[\protect\citeauthoryear{{Berry}}{{Berry}}{2015}]{Berry}
{Berry} D.~S.,  2015, \mn@doi [Astronomy and Computing]
  {10.1016/j.ascom.2014.11.004}, \href
  {http://adsabs.harvard.edu/abs/2015A%26C....10...22B} {10, 22}

\bibitem[\protect\citeauthoryear{{Berry}, {Reinhold}, {Jenness}  \&
  {Economou}}{{Berry} et~al.}{2007}]{2007ASPC..376..425B}
{Berry} D.~S.,  {Reinhold} K.,  {Jenness} T.,   {Economou} F.,  2007, in {Shaw}
  R.~A.,  {Hill} F.,   {Bell} D.~J.,  eds,  Astronomical Society of the Pacific
  Conference Series Vol. 376, Astronomical Data Analysis Software and Systems
  XVI. p.~425

\bibitem[\protect\citeauthoryear{{Bintley} et~al.,}{{Bintley}
  et~al.}{2014}]{2014SPIE.9153E..03B}
{Bintley} D.,  et~al., 2014, in Millimeter, Submillimeter, and Far-Infrared
  Detectors and Instrumentation for Astronomy VII. p. 915303,
  \mn@doi{10.1117/12.2055231}

\bibitem[\protect\citeauthoryear{{Chapin}, {Berry}, {Gibb}, {Jenness}, {Scott},
  {Tilanus}, {Economou}  \& {Holland}}{{Chapin}
  et~al.}{2013}]{2013MNRAS.430.2545C}
{Chapin} E.~L.,  {Berry} D.~S.,  {Gibb} A.~G.,  {Jenness} T.,  {Scott} D.,
  {Tilanus} R.~P.~J.,  {Economou} F.,   {Holland} W.~S.,  2013, \mn@doi
  [\mnras] {10.1093/mnras/stt052}, \href
  {http://adsabs.harvard.edu/abs/2013MNRAS.430.2545C} {430, 2545}

\bibitem[\protect\citeauthoryear{{Chapin}, {Gibb}, {Jenness}, {Berry}, {Scott}
  \& {Tilanus}}{{Chapin} et~al.}{2015}]{smurf}
{Chapin} E.,  {Gibb} A.~G.,  {Jenness} T.,  {Berry} D.~S.,  {Scott} D.,
  {Tilanus} R.,  2015, The Sub-Millimetre User Reduction Facility, Starlink
  User Note 258, Version 2.0.0.
University of British Columbia and the Science and Technologies Facilities
  Council, 2.0.0 edn

\bibitem[\protect\citeauthoryear{{Coppin} et~al.,}{{Coppin}
  et~al.}{2008}]{Coppin_SHADES}
{Coppin} K.,  et~al., 2008, \mn@doi [\mnras]
  {10.1111/j.1365-2966.2007.12808.x}, \href
  {http://adsabs.harvard.edu/abs/2008MNRAS.384.1597C} {384, 1597}

\bibitem[\protect\citeauthoryear{{Csengeri} et~al.,}{{Csengeri}
  et~al.}{2014}]{2014A&A...565A..75C}
{Csengeri} T.,  et~al., 2014, \mn@doi [\aap] {10.1051/0004-6361/201322434},
  \href {http://adsabs.harvard.edu/abs/2014A%26A...565A..75C} {565, A75}

\bibitem[\protect\citeauthoryear{Currie \& Berry}{Currie \&
  Berry}{2015}]{kappa}
Currie M.~J.,  Berry D.~S.,  2015, KAPPA -- Kernel Application Package,
  Starlink User Note 95.39, Version 2.0.0.
2.3.0 edn

\bibitem[\protect\citeauthoryear{{Currie}, {Berry}, {Jenness}, {Gibb}, {Bell}
  \& {Draper}}{{Currie} et~al.}{2014}]{2014ASPC..485..391C}
{Currie} M.~J.,  {Berry} D.~S.,  {Jenness} T.,  {Gibb} A.~G.,  {Bell} G.~S.,
  {Draper} P.~W.,  2014, in {Manset} N.,  {Forshay} P.,  eds,  Astronomical
  Society of the Pacific Conference Series Vol. 485, Astronomical Data Analysis
  Software and Systems XXIII. p.~391

\bibitem[\protect\citeauthoryear{{Dempsey} et~al.,}{{Dempsey}
  et~al.}{2013}]{2013MNRAS.430.2534D}
{Dempsey} J.~T.,  et~al., 2013, \mn@doi [\mnras] {10.1093/mnras/stt090}, \href
  {http://adsabs.harvard.edu/abs/2013MNRAS.430.2534D} {430, 2534}

\bibitem[\protect\citeauthoryear{{Doi} et~al.,}{{Doi} et~al.}{2015}]{Doi}
{Doi} Y.,  et~al., 2015, \mn@doi [\pasj] {10.1093/pasj/psv022}, \href
  {http://adsabs.harvard.edu/abs/2015PASJ...67...50D} {67, 50}

\bibitem[\protect\citeauthoryear{Gibb \& Jenness}{Gibb \&
  Jenness}{2014}]{oracdr}
Gibb A.~G.,  Jenness T.,  2014, Processing SCUBA-2 Data with ORAC-DR, Starlink
  User Note 264, Version 1.0.0.
University of British Columbia and the Science and Technology Facilities
  Council, 1.4.0 edn

\bibitem[\protect\citeauthoryear{{Holland} et~al.,}{{Holland}
  et~al.}{2013}]{2013MNRAS.430.2513H}
{Holland} W.~S.,  et~al., 2013, \mn@doi [\mnras] {10.1093/mnras/sts612}, \href
  {http://adsabs.harvard.edu/abs/2013MNRAS.430.2513H} {430, 2513}

\bibitem[\protect\citeauthoryear{{Ivison} et~al.,}{{Ivison}
  et~al.}{2007}]{Ivison}
{Ivison} R.~J.,  et~al., 2007, \mn@doi [\mnras]
  {10.1111/j.1365-2966.2007.12044.x}, \href
  {http://adsabs.harvard.edu/abs/2007MNRAS.380..199I} {380, 199}

\bibitem[\protect\citeauthoryear{{Jenness} \& {Economou}}{{Jenness} \&
  {Economou}}{2015}]{2015A&C.....9...40J}
{Jenness} T.,  {Economou} F.,  2015, \mn@doi [Astronomy and Computing]
  {10.1016/j.ascom.2014.10.005}, \href
  {http://adsabs.harvard.edu/abs/2015A%26C.....9...40J} {9, 40}

\bibitem[\protect\citeauthoryear{{Kackley}, {Scott}, {Chapin}  \&
  {Friberg}}{{Kackley} et~al.}{2010}]{PONG}
{Kackley} R.,  {Scott} D.,  {Chapin} E.,   {Friberg} P.,  2010, in {Radziwill}
  N.~M.,  {Bridger} A.,  eds,  \procspie Vol. 7740, Software and
  Cyberinfrastructure for Astronomy. SPIE, p. 77401Z,
  \mn@doi{10.1117/12.857397}

\bibitem[\protect\citeauthoryear{{MacKenzie} et~al.,}{{MacKenzie}
  et~al.}{2011}]{2011MNRAS.415.1950M}
{MacKenzie} T.,  et~al., 2011, \mn@doi [\mnras]
  {10.1111/j.1365-2966.2011.18840.x}, \href
  {http://adsabs.harvard.edu/abs/2011MNRAS.415.1950M} {415, 1950}

\bibitem[\protect\citeauthoryear{{Miller} et~al.,}{{Miller}
  et~al.}{2001}]{Miller}
{Miller} C.~J.,  et~al., 2001, \mn@doi [\aj] {10.1086/324109}, \href
  {http://adsabs.harvard.edu/abs/2001AJ....122.3492M} {122, 3492}

\bibitem[\protect\citeauthoryear{{Molinari} et~al.,}{{Molinari}
  et~al.}{2010}]{2010A&A...518L.100M}
{Molinari} S.,  et~al., 2010, \mn@doi [\aap] {10.1051/0004-6361/201014659},
  \href {http://adsabs.harvard.edu/abs/2010A%26A...518L.100M} {518, L100}

\bibitem[\protect\citeauthoryear{{Planck Collaboration VIII}}{{Planck
  Collaboration VIII}}{2016}]{2016A&A...594A...8P}
{Planck Collaboration VIII} 2016, \mn@doi [\aap] {10.1051/0004-6361/201525820},
  \href {http://adsabs.harvard.edu/abs/2016A%26A...594A...8P} {594, A8}

\bibitem[\protect\citeauthoryear{{Planck Collaboration XXVI}}{{Planck
  Collaboration XXVI}}{2016}]{2016A&A...594A..26P}
{Planck Collaboration XXVI} 2016, \mn@doi [\aap] {10.1051/0004-6361/201526914},
  \href {http://adsabs.harvard.edu/abs/2016A%26A...594A..26P} {594, A26}

\bibitem[\protect\citeauthoryear{{Planck Collaboration XXVIII}}{{Planck
  Collaboration XXVIII}}{2016}]{2016A&A...594A..28P}
{Planck Collaboration XXVIII} 2016, \mn@doi [\aap]
  {10.1051/0004-6361/201525819}, \href
  {http://adsabs.harvard.edu/abs/2016A%26A...594A..28P} {594, A28}

\bibitem[\protect\citeauthoryear{{Thompson} et~al.,}{{Thompson}
  et~al.}{2007}]{2007arXiv0704.3202T}
{Thompson} M.~A.,  et~al., 2007, preprint, \href
  {http://adsabs.harvard.edu/abs/2007arXiv0704.3202T} {} (\mn@eprint {arXiv}
  {0704.3202})

\bibitem[\protect\citeauthoryear{Watson}{Watson}{2010}]{Watson}
Watson M.,  2010, Master's thesis, University of Hertfordshire

\makeatother
\end{thebibliography}

\appendix

\section{450-\textbf{$\umu$\lowercase{m}} data}

In the 450-$\umu$m channel the atmospheric transmission is lower and the emissivity is higher, therefore for SCUBA-2 the sensitivity to sky-noise is considerably greater (about 100 times greater in most cases) compared to the 850-$\umu$m channel. To determine if creating a catalogue from the 450-$\umu$m data was feasible, we examined the SASSy field with the brightest sources of emission (the W3 star-forming region), and attempted to extract any sources from this tile. The same data reduction procedure (using an appropriately narrower matched-filter with a 7.5\,arcsec FWHM) and beam-sized source extraction techniques were performed, but only four sources within this field were recovered at 450\,$\umu$m. These are the first four sources listed in Table~\ref{tab:catalogue}.

Of the sources detected in the entire 850-$\umu$m catalogue, the next brightest has an S/N of 69 and is in the same W3 tile. This source is, however, \textit{not} detected at 450\,$\umu$m, and we therefore reason that it is not worth performing the reduction and analysis required to make an empty catalogue for the entire 450-$\umu$m data set. Based on effective exposure time results \citep[obtained as in the appendix of][]{Coppin_SHADES}, and comparison between two of the fields, we determine that the rms levels follow the scaling $\sigma_{450}$=$794(e^{26\,A\,\tau_{225}}/\textit{N$_{\rm scans}$})^{1/2}$, where $A$ is the airmass, $\tau_{225}$ is the average opacity at 225\,GHz for the tile that the source lies in, \textit{$N_{\rm scans}$} is the number of scans of the particular tile the source is located in (between three and eight), and the factor of 26 converts the opacity from 225\,GHz to 450\,$\umu$m \citep[see][]{2013MNRAS.430.2534D}. We can thus estimate $\sigma_{450}$  for each tile, knowing only effective values for $A$, $\tau_{225}$ and \textit{N$_{\rm scans}$}. Based on this, in Table~\ref{tab:catalogue} we give 3\,$\sigma$ upper limits at 450\,$\umu$m for all 850-$\umu$m sources, (except those four detected at 450\,$\umu$m).

\bsp	
\label{lastpage}
\end{document}